\documentclass[12pt]{iopart}

\usepackage[utf8]{inputenc}
\usepackage{acro}
\usepackage{units}
\usepackage{url}
\usepackage{xcolor}
\usepackage{graphicx}
\usepackage{float}
\usepackage[normalem]{ulem}

\DeclareAcronym{aLIGO}{
  short = {aLIGO},
  long = {Advanced LIGO}
}

\DeclareAcronym{ASD}{
  short = {ASD},
  long = {amplitude spectral density}
}

\DeclareAcronym{BBH}{
  short = {BBH},
  long = {binary black hole}
}

\DeclareAcronym{BNS}{
  short = {BNS},
  long = {binary neutron star}
}

\DeclareAcronym{CBC}{
  short = {CBC},
  long = {compact binary coalescences}
}

\DeclareAcronym{CDF}{
  short = {CDF},
  long = {cumulative distribution function}
}

\DeclareAcronym{CF}{
  short = {CF},
  long = {coupling function}
}

\DeclareAcronym{CS}{
  short = {CS},
  long = {corner station}
}

\DeclareAcronym{DARM}{
  short = {DARM},
  long = {differential arm length}
}

\DeclareAcronym{DQR}{
  short = {DQR},
  long = {data quality report}
}

\DeclareAcronym{EX}{
  short = {EX},
  long = {end-X station}
}

\DeclareAcronym{EY}{
  short = {EY},
  long = {end-Y station}
}

\DeclareAcronym{FC}{
  short = {FC},
  long = {filter cavity}
}

\DeclareAcronym{GW}{
  short = {GW},
  long = {gravitational wave}
}

\DeclareAcronym{HAM}{
  short = {HAM},
  long = {horizontal access module}
}

\DeclareAcronym{HVAC}{
  short = {HVAC},
  long = {heating, ventilation and air conditioning}
}

\DeclareAcronym{LIGO}{
  short = {LIGO},
  long = {Laser Interferometric Gravitational Wave Observatory}
}

\DeclareAcronym{LHO}{
  short = {LHO},
  long = {LIGO Hanford Observatory}
}

\DeclareAcronym{LLO}{
  short = {LLO},
  long = {LIGO Livingston Observatory}
}

\DeclareAcronym{NR}{
  short = {NR},
  long = {numerical relativity}
}

\DeclareAcronym{O3}{
  short = {O3},
  long = {third observing run}
}

\DeclareAcronym{O4}{
  short = {O4},
  long = {fourth observing run}
}

\DeclareAcronym{OMC}{
  short = {OMC},
  long = {output mode cleaner}
}

\DeclareAcronym{PEM}{
  short = {PEM},
  long = {physical environmental monitoring}
}

\graphicspath{figures/}

\begin{document}

\title[Automated Evaluation of Environmental Coupling for aLIGO GW Detections]{Automated Evaluation of Environmental Coupling for Advanced LIGO Gravitational Wave Detections}

\author{%
A~F~Helmling-Cornell$^{1}$,  
P~Nguyen$^{1,2}$,            
R~M~S~Schofield$^{1}$,       
and
R~Frey$^{1}$				 
\\
}
\par\medskip
\address {$^{1}$ University of Oregon, Eugene, OR 97403, USA}
\address {$^{2}$ Now at: ZAP Energy, Everett, WA 98203, USA} 
\ead{\mailto{ahelmlin@uoregon.edu}}

\begin{abstract}
The extreme sensitivity required for direct observation of gravitational waves by the Advanced LIGO detectors means that environmental noise is increasingly likely to contaminate Advanced LIGO gravitational wave signals if left unaddressed. Consequently, environmental monitoring efforts have been undertaken and novel noise mitigation techniques have been developed which have reduced environmental coupling and made it possible to analyze environmental artifacts with potential to affect the $90$ gravitational wave events detected from 2015--2020 by the Advanced LIGO detectors. So far, there is no evidence for environmental contamination in gravitational wave detections. However, automated, rapid ways to monitor and assess the degree of environmental coupling between gravitational wave detectors and their surroundings are needed as the rate of detections continues to increase. We introduce a computational tool, \textsc{PEMcheck}, for quantifying the degree of environmental coupling present in gravitational wave signals using data from the extant collection of environmental monitoring sensors at each detector. We study its performance when applied to $79$ gravitational waves detected in LIGO's third observing run and test its performance in the case of extreme environmental contamination of gravitational wave data. We find that \textsc{PEMcheck}'s automated analysis identifies only a small number of gravitational waves that merit further study by environmental noise experts due to possible contamination, a substantial improvement over the manual vetting that occurred for every gravitational wave candidate in the first two observing runs. Building on a first attempt at automating environmental coupling assessments used in the third observing run, this tool represents an improvement in accuracy and interpretability of coupling assessments, reducing the time needed to validate gravitational wave candidates. With the validation provided herein; \textsc{PEMcheck} will play a critical role in event validation during LIGO's fourth observing run as an integral part of the data quality report produced for each gravitational wave candidate.


\submitto{\CQG}

\maketitle

\end{abstract}

\section{Introduction}
\label{sec:intro}
The era of \ac{GW} astronomy began in 2015, when the \ac{aLIGO} detectors directly observed the \ac{GW}s from a \ac{BBH} merger~\cite{150914detection}. In 2017, the first \ac{GW}s from a \ac{BNS} merger were observed by the \ac{aLIGO} detectors~\cite{170817detection}. Subsequent sky localization of GW170817's host galaxy was aided by the \ac{GW} observatory Advanced Virgo, which did not observe \ac{GW}s from the merger due to the orientation of the detector relative to the source~\cite{170817detection,170817mma,advvirgo1}. In total the \ac{aLIGO} detectors, along with  Advanced Virgo and KAGRA, have observed 90 \ac{GW}s from \ac{CBC}s in the course of the first three observing runs~\cite{GWTC1,GWTC2p1,GWTC3,advvirgo2,kagra}.

The \ac{aLIGO} detectors are a pair of nearly-identical observatories. \ac{LHO} is located near Hanford, Washington, United States and \ac{LLO} is located near Livingston, Louisiana, United States. Each detector is a kilometer-scale dual-recycled Fabry-P\'erot Michelson interferometer~\cite{aligodescription,O3sensperf}. A \ac{GW} incident on one of the \ac{aLIGO} detectors produces an optical path length difference between light circulating in the two arm cavities of the interferometer. This varying difference in optical path length creates a time-dependent interference pattern when the light from the interferometer arms are recombined.

The amplitude of \ac{GW}s is expressed as the \ac{GW} strain $h$, where
\begin{eqnarray}\label{eq:gwstrain}
h=\Delta L/L
\end{eqnarray}
with $\Delta L$ the \ac{DARM} between the two interferometer arms and $L$ the unperturbed arm length, i.e. $\unit[4]{km}$. Typical \ac{GW}s observed by the \ac{aLIGO} detectors have strain amplitudes $\mathcal{O}(10^{-21})$. This extreme sensitivity results in local environmental noise easily coupling to the detector output. Examples of environmental noise seen in the \ac{O3}---which lasted from April 1, 2019 to March 27, 2020 with a commissioning break during the month of October 2019---include ground motion induced by air handling motors at \ac{LHO}, ground motion from trains traveling near \ac{LLO} and heightened anthropogenic activity at both sites during the normal workday increasing each detector's local ground motion~\cite{GWTC2p1,GWTC3,O3PEM,trainpaper,O3Scatter}. Correlated noise between the two \ac{aLIGO} detectors originating, for example, from lightning strokes, could potentially mimic or contaminate a \ac{GW} signal~\cite{MatthewMag,kagramag}. Given the precision needed to measure \ac{GW}s and the diversity of potential  environmental noise sources, it is crucial to identify times where the \ac{GW} data may be contaminated by environmental effects.

When a \ac{GW} is observed by one or more of the \ac{aLIGO} detectors, detector characterization specialists must examine both the \ac{DARM} data as well as data from a network of environmental sensors at each of the \ac{aLIGO} sites and rapidly determine:~(i)~whether the \ac{GW} candidate was a result of environmental effects in the \ac{DARM} data, or (ii)~whether there is environmental noise that is contaminating the signal, and if so, at what times and frequencies the noise occurs~\cite{O3PEM,O2O3Detchar}. While there is no evidence so far that environmental noise has measurably affected a \ac{GW} detection, the margin for safety grow smaller as the sensitivities of the \ac{aLIGO} detectors improve.

In this paper, we report on a tool, \textsc{PEMcheck}, developed to rapidly identify times and frequencies where environmental signals may couple to the \ac{DARM} measurement made by the \ac{aLIGO} detectors. The tool estimates the degree of environmental coupling and produces a recommendation for detector characterization experts to accept or reject a \ac{GW} candidate event due to the absence or presence of environmental noise contamination in the \ac{GW} data. This determination is typically completed in under $10$ minutes.

This work represents a step forward in automated \ac{GW} event validation. Prior to \ac{O3}, any assessment of environmental contamination in a \ac{GW} candidate was carried out entirely by hand~\cite{150914detection,150914pemev,BNSvetting}. During \ac{O3}, a first attempt at automatically highlighting potential environmental contributions to \ac{GW} signals was added as part of the \ac{DQR} which was generated for each potential \ac{GW} candidate~\cite{O3PEM,Philippethesis,O2O3Detchar,O3dqrdocs}. Its implementation led to a number of false positives where potential environmental coupling was identified by human event validators but was subsequently ruled out after \ac{PEM} experts reviewed \ac{PEM} sensor data and found the previous version had often been overestimating the degree of environmental coupling. Transients in \ac{PEM} channels near the \ac{GW} candidates often showed that the claimed \ac{PEM} couplings were inconsistent with the actual level of coupling during the event, usually at a frequency where only an upper limit estimate of the coupling was found (see section~\ref{ssec:envcoupling}). In this work, we have improved upon this initial design in several ways. We have designed a metric, the contamination statistic, which \ac{aLIGO} event validators can use to quickly establish whether there is substantial environmental noise in a \ac{GW} candidate signal. We have also restricted automated coupling estimates to times and frequencies near \ac{GW} candidates; previously the analysis was restricted only in time. In addition, we have implemented a sanity check for coupling estimates through a tuning procedure where we identify and reduce coupling factors which overestimate coupling to the \ac{GW} channel. The inclusion of \textsc{PEMcheck} in the \ac{O4} \ac{DQR} means that the environmental coupling assessment is available within minutes of a candidate event for the hundreds of candidate events expected to be observed in \ac{O4}~\cite{O4dqrdocs,O3bRP,lrrRates}.

This paper is organized as follows. In section~\ref{sec:PEMcf} we briefly review the \ac{aLIGO} \ac{PEM} system and methods used to quantify coupling between \ac{PEM} sensors and \ac{GW} data. In section~\ref{sec:pemcheckmethods} we detail how \textsc{PEMcheck} estimates the environmental coupling around \ac{GW} candidates as well as how the contamination statistic is calculated. In section~\ref{sec:results}, we apply \textsc{PEMcheck} to both simulated data and data from \ac{O3} \ac{GW}s. In section~\ref{sec:conclusions} we outline future directions for improving the environmental coupling estimation method presented here.

\section{\ac{aLIGO} Environmental Monitoring}
\label{sec:PEMcf}
\subsection{\ac{PEM} System Description}

There are around $100$ instruments at each \ac{aLIGO} detector which monitor the detector's local environment. This \ac{PEM} system is composed of accelerometers, microphones, magnetometers, voltage monitors, weather stations and other sensors. A map of all \ac{PEM} sensors at each \ac{aLIGO} detector is available at the \ac{aLIGO} \ac{PEM} group webpage~\cite{PEMpage}. A thorough description of the \ac{PEM} sensor network in its \ac{O3} configuration is found in~\cite{O3PEM}. Most \ac{PEM} sensor data used for this work is sampled at $\unit[4096]{Hz}$, although there are a few accelerometers and magnetometers which are sampled at $\unit[8192]{Hz}$ and $\unit[16384]{Hz}$. Detailed information on specific sensors and their configuration may be found in~\cite{O3PEM}.

The \ac{aLIGO} detectors have been modified in advance of the \ac{O4} to accommodate frequency-dependent quantum squeezing to improve the sensitivity of the observatories~\cite{O4squeezingpaper}. In addition, test masses at each detector were replaced due to defects in their mirror coatings~\cite{pointabs}. The input laser power was increased. Several baffles were added to reduce scattered light noise and some existing baffles were damped to further reduce scattered light noise~\cite{O3Scatter}. The septum window separating the \ac{OMC} from the rest of the interferometer was removed due to scattered light concerns as well. Prior to the start of \ac{O4}, changes were also made to the \ac{PEM} system. To reduce low-frequency quantum radiation pressure noise introduced by the addition of frequency-independent quantum squeezing in \ac{O3}~\cite{O3sensperf}, a $\sim\unit[300]{m}$ long \ac{FC} was constructed at each \ac{aLIGO} detector prior to \ac{O4} in order to realize frequency-dependent squeezing~\cite{O4squeezingpaper}. Accelerometers and magnetometers were added to the newly-constructed \ac{FC} endstation to monitor new potential noise coupling sites. A magnetometer monitoring the magnetic field near the \ac{LLO} \ac{EY} seismic isolation controllers was added~\cite{lloeymagalog}. Large wire coils were installed to generate magnetic fields around the detectors to study the degree of magnetic coupling. The suite of microphones at \ac{LHO} was replaced with upgraded hardware~\cite{lhomicsalog}. At both detectors, more computing space was allocated for \ac{PEM} data storage so that \ac{GW} signals up to $\unit[4]{kHz}$ could be vetted by \ac{PEM} accelerometers, magnetometers and voltage monitors.

\subsection{Studying Environmental Coupling}\label{ssec:envcoupling}

\ac{PEM} sensor data can be used to determine coupling functions that quantify the degree to which environmental effects contribute to \ac{DARM}. Coupling functions are determined by an extensive injection campaign prior to, after, and in the case of magnetometers, during each observing run. Detailed descriptions of injection procedures are found in~\cite{S6PEM,O3PEM}. Here we summarize the process for determining the coupling function for a \ac{PEM} sensor. A more thorough discussion is found in~\cite{O3PEM}. As an example, we will consider L1:PEM-CS\_ACC\_HAM6\_OMC\_Z\_DQ, a channel associated with an accelerometer at \ac{LLO} which records vacuum chamber motion in the vertical direction. This particular sensor is mounted atop \ac{HAM} 6, the vacuum enclosure which contains the \ac{OMC}---the meter-scale optical cavity which rejects unwanted light from the recombined signal from the interferometer arms, like modulation sidebands and higher order transverse modes~\cite{OMCtechnote}. The photodiodes used to measure \ac{DARM} are also situated in \ac{HAM} 6 where they witness the transmitted laser light leaving the \ac{OMC}. Ground motion couples to \ac{DARM} measurements by shaking the vacuum chamber walls or components on the optics table within this vacuum chamber, causing noise when light scattered from these shaking objects recombines with the main beam~\cite{O3Scatter}.

To quantify the response of the \ac{aLIGO} detectors to environmental perturbations, \ac{CF}s between individual \ac{PEM} sensors and the \ac{DARM} data are produced. To measure a \ac{CF} for a particular sensor, we perform many environmental noise injections across different frequency regimes and compare the response of the \ac{PEM} sensor to the response in \ac{DARM}. In the case of the \ac{OMC} accelerometer, these include playing tones on a nearby speaker and mechanically shaking the nearby vacuum enclosure. To compute the \ac{CF} for this \ac{OMC} accelerometer, we compare the frequency-domain response of the \ac{PEM} sensor and \ac{DARM} during the injection. Specifically, we compute the \ac{ASD}s of the accelerometer and \ac{DARM}, denoted by  $X_{\mathrm{inj}}(f)$ and $Y_{\mathrm{inj}}(f)$, respectively. These are compared to \ac{ASD}s of sensor and \ac{DARM} data from a quiet reference time where no environmental noise injection is occurring (denoted $X_{\mathrm{bkgd}}(f)$ and $Y_{\mathrm{bkgd}}(f)$, respectively).

At some frequencies, the detector couples strongly to the environment and the injected time \ac{DARM} \ac{ASD} clearly differs from the reference time \ac{ASD}. For frequencies where this is the case, we compute the \ac{CF} by~\cite{O3PEM}
\begin{eqnarray}\label{eq:cfdefinition}
M(f)=\sqrt{\frac{\big(Y_{\mathrm{inj}}(f)\big)^2-\big(Y_{\mathrm{bkgd}}(f)\big)^2}{\big(X_{\mathrm{inj}}(f)\big)^2-\big(X_{\mathrm{bkgd}}(f)\big)^2}}.
\end{eqnarray}
Here $M(f)$ indicates that these are ``measured" points since the detector response to an injection can be directly quantified by comparing the change in \ac{DARM} and sensor \ac{ASD}s.

At frequencies where the interferometers are well-isolated from environmental noise sources, a \ac{CF} cannot be measured since the \ac{DARM} \ac{ASD} is unperturbed by the injection. In this case, we may only set an upper limit on the coupling between the \ac{PEM} sensor and \ac{DARM}. An upper limit estimate, $U(f)$, of the coupling may be set by assuming that at these frequencies, any noise in the \ac{DARM} \ac{ASD} is due to noise witnessed by the \ac{PEM} sensor, or~\cite{O3PEM}
\begin{eqnarray}\label{eq:uldefinition}
U(f)=\frac{Y_{\mathrm{bkgd}}(f)}{\sqrt{\big(X_{\mathrm{inj}}(f)\big)^2-\big(X_{\mathrm{bkgd}}(f)\big)^2}}.
\end{eqnarray}

\begin{figure}
\includegraphics[width=\textwidth]{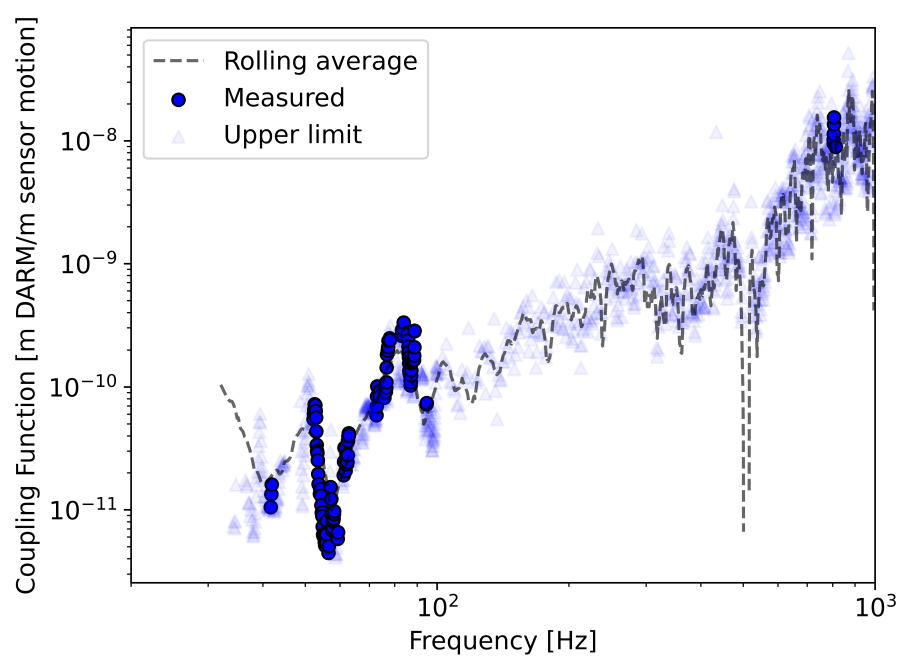}
\caption{\label{fig:example_cf} Vibrational coupling between \ac{LLO}'s \ac{HAM} 6 vacuum chamber Z-axis accelerometer and \ac{DARM} as measured prior to \ac{O3}. The vibrational coupling here is likely driven by stray light scattering off of the septum window dividing this vacuum chamber and the adjoining \ac{HAM} 5 chamber and then recombining with the main beam~\cite{septumalog}. The large fraction of upper limit estimates, rather than measured values, for the vibrational coupling is because \ac{PEM} injections could not be increased to an amplitude such that the injection was visible in the \ac{DARM} data without saturating the accelerometer signal. We directly convert the acceleration measured by the accelerometer into displacement of the vacuum chamber wall. The resonance near \unit[100]{Hz} could be a resonance of the vacuum chamber wall, the table, or optics on the table. Near $\unit[500]{Hz}$ the \ac{CF} is set to $\unit[0]{m/m}$ because the detector is insensitive around the frequencies of the test mass suspension resonances (violin modes)~\cite{GWOSCpaper}.}
\end{figure}

The measured points and upper limit estimates on the coupling calculated via equations~\ref{eq:cfdefinition} and~\ref{eq:uldefinition} are combined across several injections in different frequency bands to form a \ac{CF} for each \ac{PEM} sensor that runs from a few $\unit[]{Hz}$ to a few $\unit[]{kHz}$. Each sensor's \ac{CF} is a mix of measured points and upper limit coupling estimates. The value of the \ac{CF} at a given frequency is chosen by identifying the highest-amplitude injection in the \ac{PEM} sensor at that frequency (i.e. the injection with the largest value for $X_{\mathrm{inj}}(f)$ that frequency). the \ac{CF} at the particular frequency takes the value of the measured or upper limit estimate of the coupling computed by equation~\ref{eq:cfdefinition} or~\ref{eq:uldefinition} for this highest-amplitude injection. Figure~\ref{fig:example_cf} shows the \ac{CF} computed for L1:PEM-CS\_ACC\_HAM6\_OMC\_Z\_DQ by combining several \ac{PEM} injections which was used during \ac{O3}.

Potential sources of uncertainty for measured points in the \ac{CF} arise from, for example, the finite spacing of \ac{PEM} sensors. An environmental signal may be louder at an unmonitored coupling site than at the nearest sensor, resulting in an underestimation of the signal's effect on the \ac{GW} data. Occasional environmental disturbances, such as the thunderclap discussed in section~\ref{ssec:simresults}, show that predictions for \ac{DARM} made using \ac{PEM} and \ac{CF} data correspond to the recorded value of \ac{DARM} during the environmental transient with a factor of $\sim2$ uncertainty~\cite{O3PEM}. The true coupling between a \ac{PEM} sensor and the \ac{DARM} is given by a log-normal probability distribution since systematic uncertainties in the injection procedure prevent an entirely accurate measurement of the coupling at a given frequency~\cite{Philippethesis}.

We may estimate the contribution of environmental noise to the \ac{DARM} measurement by projecting
\begin{eqnarray}\label{eq:cfprojection}
Y_{\mathrm{noise}}(f)=CF(f)\times X_{\mathrm{\ac{GW}}}(f)
\end{eqnarray}
where $X_{\mathrm{\ac{GW}}}(f)$ is the \ac{ASD} of time series data in a given \ac{PEM} sensor around the time of a \ac{GW} candidate, $Y_{\mathrm{noise}}(f)$ is the \ac{ASD} of the \ac{DARM} time series recorded at the \ac{GW} candidate time solely due to environmental influences witnessed by the \ac{PEM} sensor, and $CF(f)$ the sensor's coupling function.

For this work, we use \ac{CF}s measured for \ac{O3}~\cite{peminjso3h,peminjso3l}. Injection campaigns at both \ac{aLIGO} detectors have been completed to update the \ac{CF}s for their use in \ac{O4}~\cite{peminjso4h,peminjso4l}.

\section{Quantifying Environmental Coupling}
\label{sec:pemcheckmethods}
\subsection{Data Access}

The \textsc{PEMcheck} analysis requires basic information about the \ac{GW} candidate it is studying.  This information can either be input manually by the user or it can be supplied automatically in the form of a \textsc{GraceDB} Superevent ID~\cite{GraceDB}. If given a Superevent ID, \textsc{PEMcheck} identifies the preferred event if there are a collection of triggers from different search pipelines all associated with the same \ac{GW} candidate. We use the \textsc{GWpy} software package for data access and signal processing~\cite{gwpy}.

\textsc{PEMcheck} is designed for data quality assessments of short-duration \ac{GW} transients. While it is best-suited for studying \ac{GW}s from \ac{CBC}s, it can also be extended to analyze brief bursts of \ac{GW}s from poorly modeled sources (e.g.~cosmic strings, highly eccentric \ac{BBH} mergers, etc.)~\cite{O3CS,O3eBBH}. For events identified by the low-latency \ac{CBC} search pipelines \textsc{GstLAL}, \textsc{MBTA}, \textsc{PyCBC~Live} and \textsc{SPIIR}, \textsc{PEMcheck} requires the masses and spins of each of the compact objects in the \ac{CBC} as well as the \ac{GW} candidate time~\cite{lal1,lal2,mbta,pycbc1,pycbc2,spiir1,spiir2}. The waveform parameters can be extracted from the \ac{GW} candidate's preferred event data on \textsc{GraceDB}. These parameters are used to approximate the time-frequency evolution of the \ac{GW} signal in the \ac{aLIGO} detectors using a non-precessing waveform model for \ac{CBC}s which incorporates terms up to order 4 in the post-Newtonian expansion of the effective-one-body gravitational potential. The particular \ac{NR} approximant used by \textsc{PEMcheck} also employs reduced-order modelling of the potential which reduces the time needed to compute the \ac{GW} waveform with a minimal effect on waveform accuracy~\cite{SEOBNRv4ROM}. In the case of a short, unmodeled \ac{GW} transient, \textsc{PEMcheck} requires the time and frequency ranges at which \ac{GW} strain was observed by the \textsc{cWB} pipeline~\cite{cwb1,cwb2}. This information is also available on \textsc{GraceDB} if a \textsc{cWB} trigger is identified as the preferred event. The time and frequency information restricts the search for environmental coupling to the time and frequency ranges at which the \ac{GW} candidate is witnessed by the detectors, which was not the case for the pre-\ac{O4} tool used to evaluate the presence of environmental noise in the \ac{DARM} data. In addition to automatic parsing of \textsc{GraceDB} data, \textsc{PEMcheck} also accepts manual entry of waveform parameters or rectangular regions in time-frequency space where excess energy is observed in the \ac{GW} data. In the case of a noteworthy \ac{GW} candidate during \ac{O4}, such as the first detection of \ac{GW}s from a new source type, \ac{PEM} experts will still vet the data by hand.

\begin{table}
\caption{\label{tab:GW_basic_props} Basic detection information and initial estimates of source frame masses and dimensionless spins for GW190707. This information is used to estimate the \ac{GW} waveform shown in figure~\ref{fig:cvalsomc}.}
\begin{indented}
\item[]\begin{tabular}{@{}ll}
\br
Online \ac{aLIGO} detectors & \ac{LHO}, \ac{LLO}\\
$t_c$ (GPS time, $\unit[]{s}$) & $1246527224.18$\\
$m_1/M_{\odot}$ & $15.51$\\
$m_2/M_{\odot}$ & $8.62$\\
$\chi_1$ & $-0.30$\\
$\chi_2$ & $0.76$\\
\br
\end{tabular}
\end{indented}
\end{table}

Here we use GWTC-3 event GW190707\_093326 (hereafter GW190707) as an example to describe how \textsc{PEMcheck} estimates the extent to which environmental noise contaminates the \ac{GW} data~\cite{GWTC2,GCN25012}. Table~\ref{tab:GW_basic_props} lists the inferred merger properties used by \textsc{PEMcheck} for waveform approximation. We restrict the following discussion to \ac{LLO} data.

\subsubsection{\ac{DARM} and \ac{PEM} data}\label{sssec:timeseries}

Once the time and frequency ranges of the \ac{GW} candidate are established, \textsc{PEMcheck} requests \ac{DARM} and \ac{PEM} sensor data from around the candidate time. The \ac{DARM} data is sampled at $\unit[16384]{Hz}$. This time series data is split into foreground and background times. For \ac{CBC}s, the default definition for the foreground time is $t_{\mathrm{fore}}=[t_{20},t_c]$, where $t_{20}$ is the earliest time at which the \ac{GW} signal is greater than or equal to $\unit[20]{Hz}$ and $t_c$ is the merger time as specified by the \textsc{GraceDB} preferred event. For each candidate event, a suitable background time period is also identified. For a \ac{CBC} signal, the minimum background time is $t_{\mathrm{back}}=[t_{20}-\unit[32]{s},t_{20})$. For a burst signal, $t_{\mathrm{fore}}=[t_{\mathrm{start}},t_{\mathrm{end}}]$ and $t_{\mathrm{back}}=[t_{\mathrm{start}}-\unit[32]{s},t_{\mathrm{start}})$, with $t_{\mathrm{start}}$ and $t_{\mathrm{end}}$ the beginning and end of the time period identified by the burst search pipeline. The foreground time is defined to capture the behavior of the \ac{PEM} sensors and the \ac{GW} data during the \ac{GW} candidate itself. The background time is chosen to provide the longer-term behavior of these channels. A longer background time means that any glitches in the \ac{GW} channel which occur during the background are averaged over, reducing the amount they affect this calculation. In the case of a longer-duration \ac{GW} signal, the background time can extend to up to $\unit[256]{s}$ prior to $t_{20}$ or $t_{\mathrm{start}}$ for a \ac{CBC} or burst signal, respectively. The background duration is increased from the minimum when the duration of the foreground period is comparable to the minimum background duration of $\unit[32]{s}$. In the case of GW190707, $t_{20}=\unit[1246527219.46]{s}$ and $t_c$ is given in table~\ref{tab:GW_basic_props}.

\subsection{\ac{ASD} Projection}\label{ssec:projected}

\begin{figure}
\includegraphics[width=\textwidth]{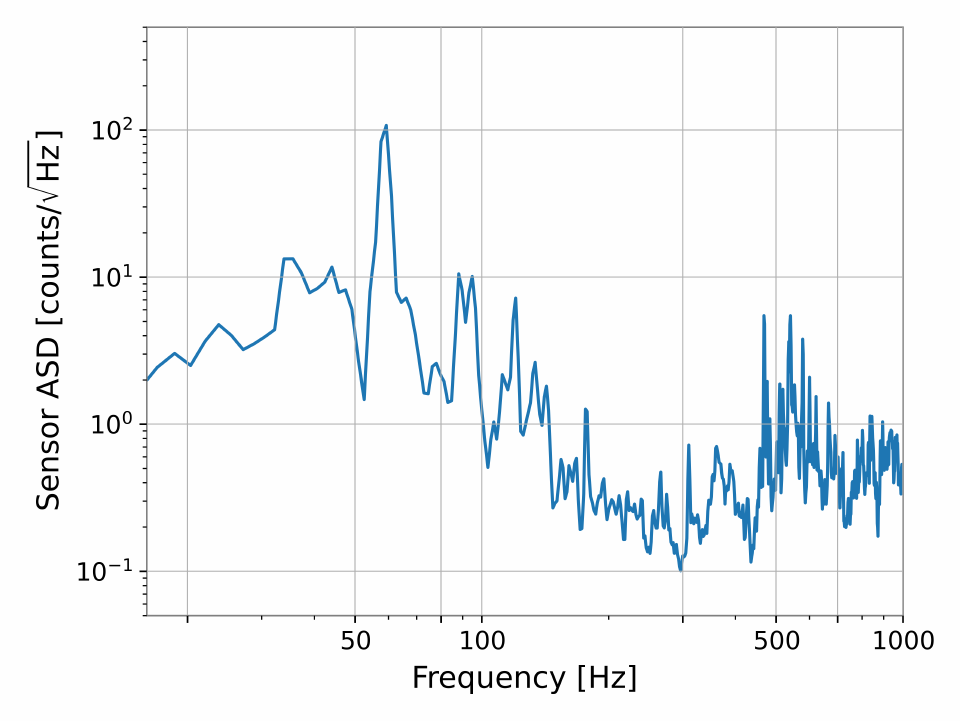}
\caption{\label{fig:asd_aux} \ac{ASD} of the Z-axis \ac{HAM} 6 accelerometer during the background time. Background \ac{PEM} sensor data is used to compute the tuned \ac{CF} of section~\ref{sssec:tunedasd}. $\unit[45.75]{s}$ of sensor data were used to calculate this \ac{ASD}. Note that each sensor count corresponds to $\unit[6.1]{\mu m~s^{-2}}$ of acceleration in the vertical direction~\cite{PEMpage}.}
\end{figure}

Using the time series data described in section~\ref{sssec:timeseries}, we calculate the \ac{ASD} of \ac{DARM} and each \ac{PEM} sensor data during both the foreground and background time segments. The duration, $\tau$, of each fast Fourier transform used in the \ac{ASD} calculation is given by
\begin{equation}\label{eq:fft_len}
\tau=\cases
{
\unit[0.25]{s} & if $t_{\mathrm{fore}}<\unit[2]{s}$ \\
\unit[t_{\mathrm{fore}}/8]{} & if $\unit[2]{s}\le t_{\mathrm{fore}}\le\unit[8]{s}$ \\
\unit[1]{s} & otherwise \\
}
\end{equation}
and the overlap fraction is $\max(1/2, 1-t_{\mathrm{fore}}/16\tau)$. Welch's method with median averaging and a Hann window is used to calculate the \ac{ASD} for each channel~\cite{welch}. Figure~\ref{fig:asd_aux} shows the \ac{ASD} of the \ac{PEM} sensor background data.

We now discuss the projection of environmental noise into the \ac{DARM} data during $t_{\mathrm{fore}}$. For each \ac{PEM} sensor studied by \textsc{PEMcheck}, we first linearly interpolate between the \ac{CF} data points so that the frequency resolution of the \ac{CF} matches the frequency resolution of the \ac{GW} data. We then tune the \ac{CF} for the sensor if the predicted level of noise in \ac{DARM} due to the \ac{PEM} data exceeds the actual level of noise during the background time. We next compute the \ac{ASD} of the \ac{PEM} sensor and \ac{DARM} data using the prescription given in equation~\ref{eq:fft_len}. Then, for each frequency bin in the \ac{PEM} \ac{ASD}, we compute the meters of \ac{DARM} during the \ac{GW} candidate caused by environmental noise via equation~\ref{eq:cfprojection}.

\subsubsection{\ac{CF} Tuning}\label{sssec:tunedasd}

\begin{figure}
\includegraphics[width=\textwidth]{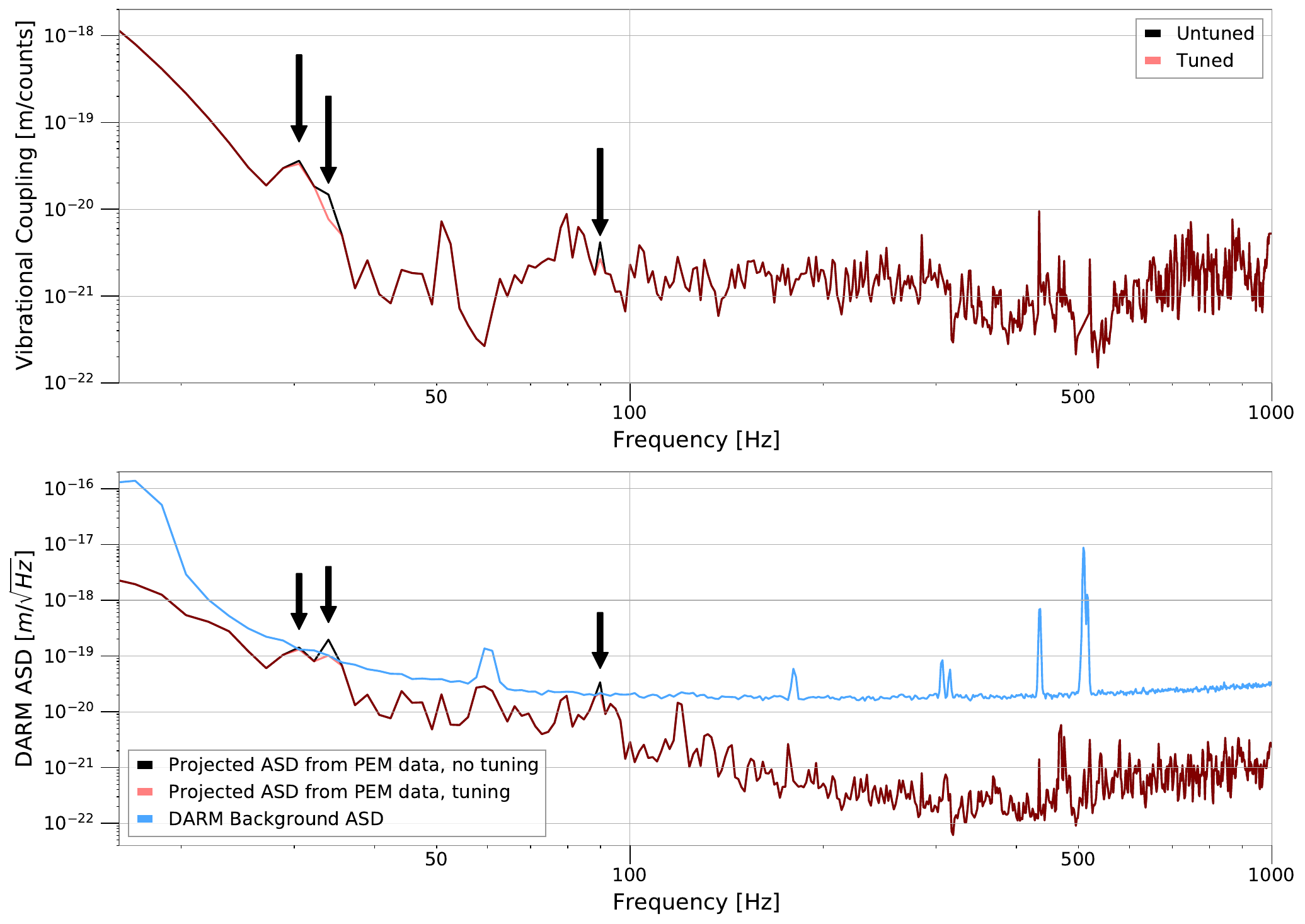}
\caption{\label{fig:tuningasds} Top: Comparison of the \ac{CF} interpolated from the data in figure~\ref{fig:example_cf} to the interpolated, tuned \ac{CF} computed by analyzing the \ac{DARM} background \ac{ASD}. The untuned and tuned predictions for the \ac{ASD} largely agree, except near $30$ and $\unit[90]{Hz}$. The points where environmental contributions to \ac{DARM} are overestimated and need tuning are derived from upper limit estimates of the chamber motion to \ac{DARM} coupling. These frequencies are marked with black arrows. Bottom: Comparison of the \ac{DARM} $t_{\mathrm{back}}$ \ac{ASD} with the estimated contribution from the \ac{HAM} 6 accelerometer. The result of the tuning procedure is to constrain the predicted environmental contribution from a sensor to be not greater than the observed \ac{DARM} \ac{ASD}. As in the top plot, black arrows mark where the untuned \ac{CF} predicts that the \ac{DARM} \ac{ASD} should exceed its observed value to the \ac{HAM} 6 accelerometer data. These are the points which require tuning to reconcile the predicted and actual \ac{DARM} \ac{ASD}s.}
\end{figure}

Sometimes the \ac{CF}s used for \textsc{PEMcheck} overestimate the level of environmental noise present in the \ac{GW} data. To reduce environmental noise coupling overestimation, we \textit{tune} the interpolated \ac{CF}s supplied to \textsc{PEMcheck} for each \ac{GW} candidate. This process works by comparing $Y_{\mathrm{noise}}(f)$ during the background time to the actual background time \ac{DARM} \ac{ASD}. At frequencies where the environmental noise \ac{ASD} is predicted to exceed the \ac{DARM} \ac{ASD}, we reduce $CF(f)$ at those frequencies such that the predicted environmental noise is equal to the value of the \ac{DARM} \ac{ASD} at those frequencies. This process is illustrated in the top plot in figure~\ref{fig:tuningasds}. By reducing the \ac{CF} at points where there is no evidence for the claimed coupling during $t_{\mathrm{back}}$ we limit the instances of overestimating environmental coupling.

Overestimating environmental coupling may lead to event validators spending additional time investigating noise sources identified by \textsc{PEMcheck} that are not truly present in the \ac{GW} data. Typically this manual review process involves identifying transients in the \ac{PEM} sensor data at the same frequency as the reported coupling and close to the \ac{GW} candidate in time and checking whether the predictions for environmental noise contamination are consistent for these loud transients. There may also be difficulties interpreting the astrophysical parameters of a given \ac{GW} event. The difficulty disentangling the effects of offline noise subtraction from novel features in a \ac{GW} signal was highlighted in the context of noise subtraction due to glitches---not environmental noise---during \ac{O3}~\cite{localizationnoise,200129precession,200129noprecession}. Event validators incorrectly prescribing noise subtraction due to \textsc{PEMcheck} overestimating the degree of environmental noise present in a \ac{GW} signal may lead to similar parameter estimation issues in \ac{O4}.

In the case of the \ac{LLO} \ac{HAM} 6 accelerometer and GW190707, the \ac{CF} overestimates the level of environmental noise contribution to the \ac{DARM} measurement, particularly between $\sim\unit[30-40]{Hz}$ and near $\unit[90]{Hz}$, as illustrated in both plots of figure~\ref{fig:tuningasds}. In this case, environmental coupling projections derived from upper limit estimates of the coupling, rather than measured points in the \ac{CF} exceed the actual level of noise in the detector data. The coupling estimate is tuned at these frequencies to be consistent with the observed \ac{GW} data. The tuning procedure reduces the need for manual vetting of \ac{GW} signals by down-weighting spurious predictions like this one.

Disagreement between the \ac{HAM} 6 accelerometer \ac{CF} and the observed coupling could arise due to differences between the \ac{ASD} used in calculating the reference \ac{ASD} during sub--$\unit[100]{Hz}$ vibrational and acoustic injections and the background \ac{ASD} data shown in figure~\ref{fig:asd_aux}~\cite{O3PEM}. These differences may arise due to different noise sources being present in the data in March and April 2019, when the \ac{CF}s were measured and July 2019, when that \ac{GW} was observed. Additionally, commissioning efforts in the intervening months increased the low--frequency sensitivity of the \ac{LLO} detector, further constraining upper--limit estimates of the environmental coupling where it was not measured.

By tuning each \ac{CF} to account for the environmental noise and condition of the \ac{aLIGO} detector at the \ac{GW} candidate time, we improve the accuracy of environmental noise contamination information presented to \ac{aLIGO} event validators compared to the information given to \ac{O3} event validation experts~\cite{O3PEM,O2O3Detchar}.

\subsection{$c$-statistic Determination}

From the \ac{ASD} of the \ac{DARM} data during the candidate \ac{GW} event as well as the predicted \ac{DARM} \ac{ASD} curve due to environmental noise witnessed by a particular \ac{PEM} sensor, we quantify the degree to which environmental disturbances may contaminate \ac{GW} strain data. 

Following equation~\ref{eq:gwstrain}, we divide the tuned \ac{CF} by a factor of $\unit[4000]{m}$ (the length of the \ac{aLIGO} Fabry-P\'erot arm cavities) to convert the \ac{CF} from units of differential arm length per sensor unit observed to units of \ac{GW} strain per sensor unit~\cite{aligodescription}. We then compute the spectrogram of the \ac{GW} strain data during the foreground time as well as the spectrogram of the effective strain measurement due to environmental coupling for each \ac{PEM} channel, as described in section~\ref{ssec:projected}. We denote the predicted value of the \ac{GW} strain spectrogram tiles due solely to environmental contamination as $\mu(t,f)$, where
\begin{eqnarray}
\mu(t,f)=\Big(CF_{\mathrm{tuned}}(f)/L\Big)\times S_{\mathrm{PEM}}(t,f)
\end{eqnarray}
and $S_{\mathrm{PEM}}(t,f)$ is the foreground-time spectrogram of the \ac{PEM} sensor data. The predicted spectrogram is then compared to the spectrogram of the \ac{GW} strain, which we denote as $h(t,f)$.

For frequencies where $\mu$ was computed using a measured \ac{CF}, then---as mentioned in section~\ref{ssec:envcoupling}---the probability of observing a strain $h$ due to a local environmental disturbance is given by the log-normal probability density function with a standard deviation of $\ln(2)$, or
\begin{eqnarray}\label{eq:meas_pdf}
\rho(h,\mu)=\frac{1}{h(t,f)\ln(2)\sqrt{2\pi}}\exp\Bigg(-\frac{\Big(\ln(h(t,f))-(\ln(\mu(t,f))\Big)^2}{2\big(\ln(2)\big)^2}\Bigg).
\end{eqnarray}
For frequencies where $\mu$ was computed using an upper-limit \ac{CF}, we use a uniform probability distribution on the range $[0,\mu)$ to estimate the probability density function of observing the \ac{GW} strain, i.e. 
\begin{equation}\label{eq:ul_pdf}
\rho(h,\mu)=\cases
{
\frac{1}{\mu(t,f)} & if $h(t,f)<\mu(t,f)$ \\
0 & otherwise \\
}.
\end{equation}
The strict limit on the probability density function here is because the pre-observing run \ac{PEM} injections are unable to establish a coupling between the sensor data and the \ac{GW} data at these frequencies. For \ac{GW} strains that are larger than the stringent upper limits set on the coupling times the \ac{PEM} sensor data, we assume that the likelihood of environmental noise coupling vanishes.

To describe the likelihood of environmental coupling near a \ac{GW} event we define the contamination statistic, denoted by $c$, which quantifies the likelihood of recording a strain value in excess of $h(t,f)$ due to environmental noise alone. We compute it via the \ac{CDF} of the distributions in equations~\ref{eq:meas_pdf} and~\ref{eq:ul_pdf}. The $c$-statistic at each point in the \ac{GW} spectrogram is given by

\begin{equation}\label{eq:csc_def}
c(t,f)=\cases
{
\frac{1}{2}\Bigg[-\mathrm{erf}\Bigg(\frac{\ln(h(t,f))-\ln(\mu(t,f))}{\sqrt{2}\ln(2)}\Bigg)-1\Bigg] & $f\in(f_{\mathrm{meas}})$ \\
\mathrm{min}\Bigg(\frac{h(t,f)}{\mu(t,f)},1\Bigg) & $f\notin(f_{\mathrm{meas}})$ \\
}.
\end{equation}
where $f_{\mathrm{meas}}$ denotes the set of frequencies at which the \ac{CF} was measured, rather than an upper limit on the \ac{CF} estimated. Here $\mathrm{erf}()$ is the error function and $\mathrm{min}()$ denotes the minimum of the two numbers.

The $c$-statistic is distributed over the range $[0,1]$. A low $c$-statistic for a given sensor indicates that environmental disturbances witnessed by the \ac{PEM} sensor could account for some of the data in the \ac{GW} strain channel at a given time and frequency range. A high $c$-statistic indicates that it is unlikely that environmental noise witnessed by the \ac{PEM} sensor is coupling to the \ac{GW} data. Environmental contamination of the \ac{GW} data is indicated by one or more sensors reporting low $c$-value, inconsistent with the low-noise case, where many sensors should report a $c$-statistic near $1$. Results from computing the $c$-statistic many times over all the \ac{PEM} data channels are shown in section~\ref{ssec:realresults}. These distributions largely reflect the expectation for the distribution of the $c$-statistic for the null hypothesis (negligible environmental contamination), as discussed in section~\ref{ssec:realresults}.

\begin{figure}
\includegraphics[width=\textwidth]{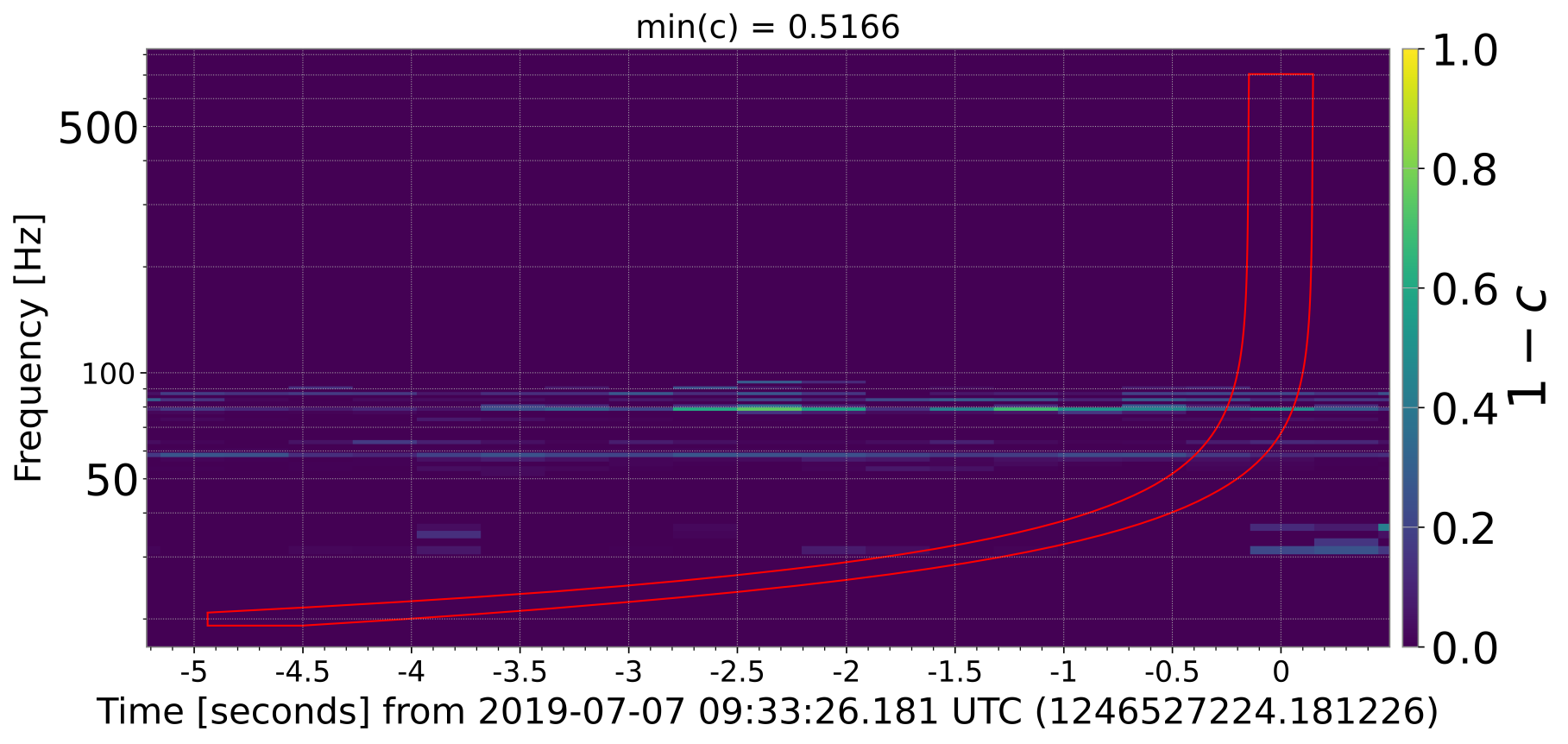}
\caption{\label{fig:cvalsomc} The results of the $c$-statistic calculation for each spectrogram tile for the \ac{LLO} \ac{HAM} 6 accelerometer during GW190707. The value of $1-c$ is plotted for each tile so that time-frequency tiles with the highest likelihood of contamination, as calculated by \textsc{PEMcheck}, appear brighter. The time-frequency tiles enclosed by the red box track the evolution of the \ac{GW} signal as predicted by \ac{NR}. The lowest $c$-statistic found within the \ac{GW} track occurs at $\sim\unit[0.1]{s}$ and $\sim\unit[78]{Hz}$ prior to the event's $t_c$. A minimum $c$-statistic of $0.52$ indicates that it is unlikely that environmental disturbances witnessed by this sensor couple to the \ac{GW} strain data during the event.}
\end{figure}

Not every spectrogram tile's $c$-statistic is relevant to determining whether there is environmental noise contamination in the \ac{GW} strain signal.  For instance, \ac{aLIGO} detected \ac{GW}s at a maximum frequency of $\sim\unit[693]{Hz}$ during GW190707; any environmental noise that was present in $\unit[]{kHz}$ frequencies, whether it couples strongly or not to the \ac{DARM} measurement, should not affect the \ac{GW} signal reconstruction. Therefore, we only consider the $c$-statistic for spectrogram tiles whose central frequencies are within $\unit[1-4]{Hz}$ of the time-frequency track predicted by the \ac{NR} waveform approximant or the time-frequency ``box" predicted by the \ac{GW} burst search. The results of the $c$-statistic calculation for the \ac{LLO} accelerometer during GW190707 are shown in figure~\ref{fig:cvalsomc}.

\subsection{Combining Results}~\label{ssec:combining}

\textsc{PEMcheck} reports to the user the time-frequency region with the lowest $c$-statistic from the entire collection of \ac{PEM} channels analyzed. From the results presented in section~\ref{sec:results}, we find that a \ac{GW} candidate with a minimum $c$-statistic of $0.2$ or less for one or more \ac{PEM} channels seems to be a reasonable threshold in \ac{O4} for follow up by environmental noise experts. This threshold identifies \ac{GW} candidates with the most significant evidence for environmental coupling while minimizing the number of events needed to be hand-vetted. If this threshold is too restrictive in \ac{O4}, it may be changed. For each \ac{GW}, a report may be automatically generated showing the time and frequency location of the minimum $c$-statistic tile near the \ac{GW} data for each \ac{PEM} sensor.

The coupling estimation process outlined here is valid only for linear coupling between one or more \ac{PEM} channels and the \ac{GW} data. Analyzing nonlinear coupling---where environmental disturbances at a particular frequency pollute \ac{GW} data at a different frequency---is typically done from a glitch mitigation perspective, and is beyond the capability of \textsc{PEMcheck} at this time.

\section{Results}
\label{sec:results}
\subsection{Real \ac{GW} Data}\label{ssec:realresults}

We report the results of running \textsc{PEMcheck} on data from each \ac{aLIGO} detector around all $79$ \ac{GW} events detected by \ac{aLIGO} in \ac{O3} with a probability of astrophysical origin of $0.5$ or greater~\cite{GWTC2p1,GWTC3,GWOSCpaper,pastro}. This amounts to $149$ individual runs of \textsc{PEMcheck}, one per \ac{GW} event per online \ac{aLIGO} detector. The sensor \ac{CF}s determined prior to the start of \ac{O3} were used for the following analyses, with the exception of results presented in table~\ref{tab:tuningstats}.

\begin{figure}
\includegraphics[width=\textwidth]{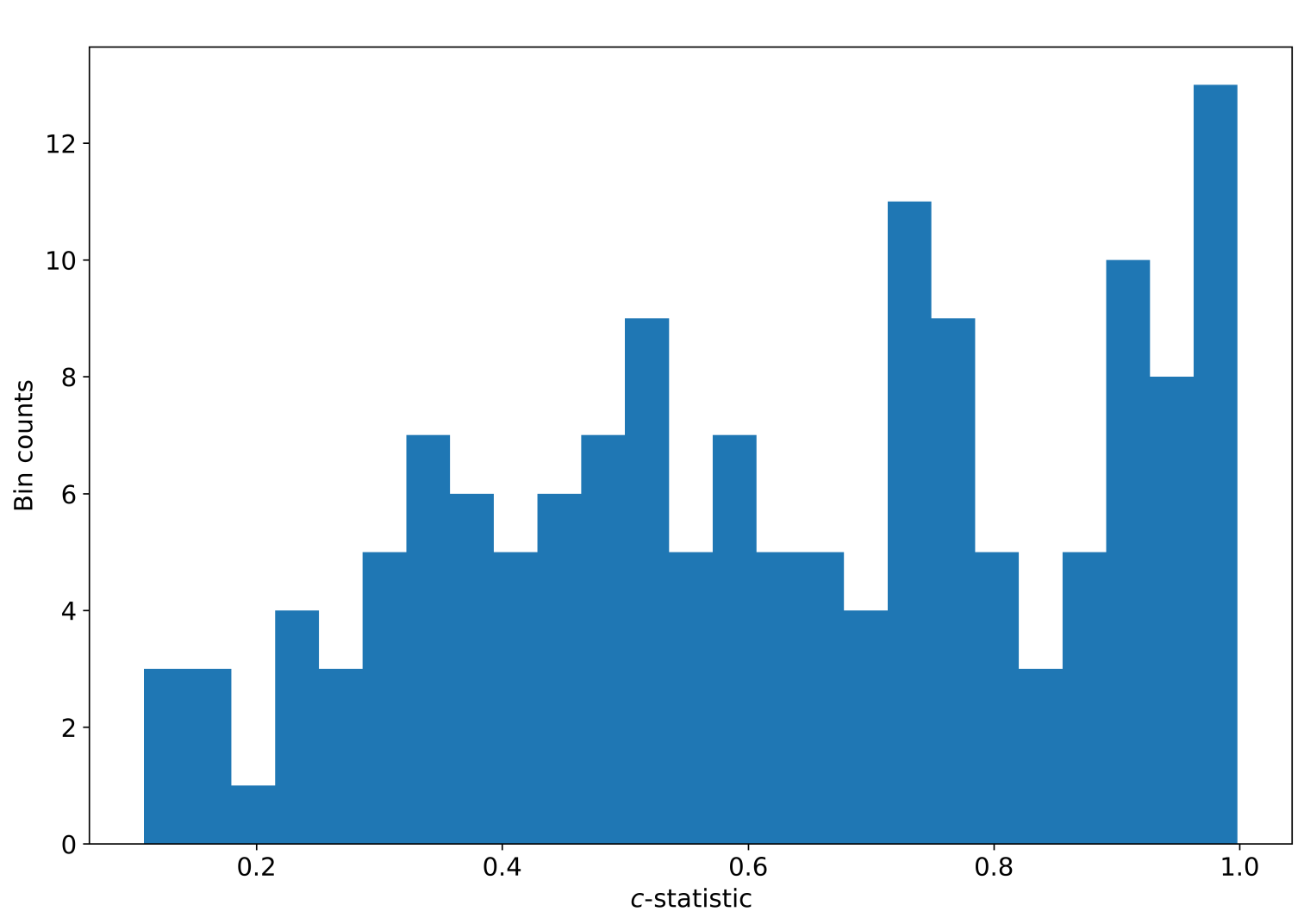}
\caption{\label{fig:cstatdist} Histogram of the minimum $c$-statistic calculated by \textsc{PEMcheck} for all O3 events using the pre-\ac{O3} coupling data. The \textsc{PEMcheck} analysis identifies $6$ of the $149$ \ac{O3} \ac{GW} events as having a minimum $c$ below $0.2$}.
\end{figure}

\begin{figure}
\includegraphics[width=\textwidth]{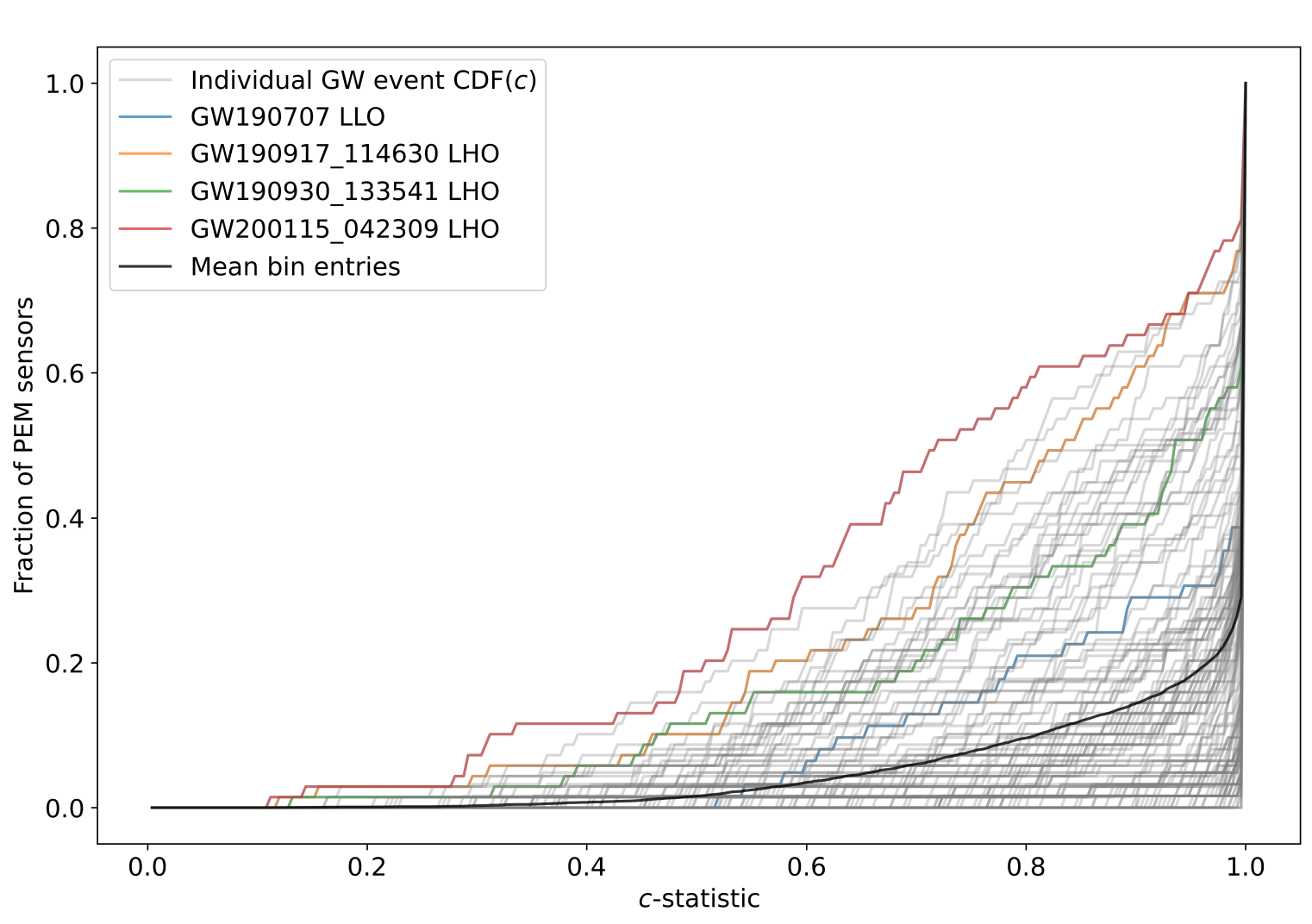}
\caption{\label{fig:spaghetti_plot} Cumulative distribution of the $c$-statistic found for each \ac{PEM} channel studied by \textsc{PEMcheck} for each \ac{GW} event in \ac{O3} using the pre-\ac{O3} coupling data. The coupling analysis was performed on $63$ \ac{PEM} sensors at \ac{LLO} and $72$ \ac{PEM} sensors at \ac{LHO}. Each light grey trace corresponds to the \ac{CDF} of the $c$-statistic for an individual \ac{GW} event, while the black trace denotes the mean number of sensors with that particular value of $c$ or less. GW190707 as well as the three events with the lowest $c$ are highlighted.}
\end{figure}

We find that \textsc{PEMcheck} does not report a $c$-statistic of $0.1$ or less for any of the $149$ runs. The distribution of the minimum $c$-statistic for all $149$ instances of \textsc{PEMcheck} is shown in figure~\ref{fig:cstatdist}. Figure~\ref{fig:spaghetti_plot} shows the cumulative distribution of $c$-statistics for all channels for all \ac{O3} events analyzed in this work.  Overall, most \ac{GW} events report a large $c$, or a low likelihood of environmental noise contamination. However, there are a few outliers in this set which merit further consideration. While all the steps involved with manually vetting \ac{GW} candidates is beyond the scope of \textsc{PEMcheck}, it provides a starting point for where to look for evidence of environmental coupling. In sections~\ref{ssec:S200115j}-\ref{sssec:S190930s} we discuss the events with the lowest minimum $c$-statistics in more detail.

We also find that the tuning procedure was frequently used to down-weight the predicted level of environmental contamination. Table~\ref{tab:tuningstats} shows the number of events where the time-frequency tile with the highest likelihood of contaminating the \ac{GW} data was derived from a tuned projection of the \ac{PEM} channel coupling. The data indicate that the \ac{CF}s found via the \ac{PEM} injection procedure predicted realistic estimates of environmental noise about two-thirds of the time. However, about one-third of events had initial noise projections which were inconsistent with the actual \ac{DARM} \ac{ASD} and needed to be tuned. In many of these cases an upper limit estimate of the coupling, where no response to a \ac{PEM} injection is seen in the \ac{GW} data, turns out to overestimate the environmental noise. This is the intended utility of the tuning procedure. However there were also $22$ instances of a measured coupling between a \ac{PEM} sensor and the {GW} channel overestimating the environmental noise contamination. Of these, $16$ stem from a measured coupling at either $148$ or $\unit[196]{Hz}$ between an accelerometer, H1:PEM-CS\_ACC\_ISCT6\_SQZLASER\_X\_DQ, on the optics table which houses many of the components of \ac{LHO}'s squeezer. This accelerometer's coupling consistently needed to be tuned down at $\unit[148]{Hz}$ starting with GW190814, which was observed the day after the squeezer table configuration was modified due to the original squeezer laser failing~\cite{190814discovery,sqzalog}. It is possible that the squeezer configuration changes affected the coupling between squeezer optics table motion and the \ac{GW} channel. After reanalyzing the dataset with this sensor's \ac{CF} as measured at the conclusion of \ac{O3}, we find that only $7$ minimum $c$-statistic tiles are derived from tuned, measured values of the coupling. None of the remaining $7$ tuned, measured points stem from the \ac{LHO} squeezer table vibrational coupling being overestimated. For these $7$ points, tuning reduced the predicted environmental noise in the \ac{GW} channel by a median of $18\%$ compared to the untuned prediction for the \ac{GW} channel \ac{ASD} at the worst coupling frequency.

\begin{table}
\caption{\label{tab:tuningstats} The number of \ac{O3} \ac{GW} events where the tuning procedure was used in calculating the time-frequency tile with the highest likelihood of contamination. The type of point, measured or upper limit, in the worst coupling sensor \ac{CF} at that frequency is also given. Most, but not all, tuning is done at frequencies where sensor \ac{CF}s are given by upper limit estimates of the coupling. We compare the fraction of tuned, measured points identified by \textsc{PEMcheck} when the \ac{CF}s measured prior to \ac{O3} are used to the fraction of tuned, measured points identified when the \ac{CF} data for \ac{LHO}'s squeezer table accelerometer is replaced with the \ac{CF} data collected after \ac{O3} for \ac{GW} events occurring after August 13, 2019.}
\begin{tabular}{lllll}
\br 
& \multicolumn{2}{c}{Untuned}  & \multicolumn{2}{c}{Tuned} \\
\ac{CF} data used & Measured  & Upper Limit & Measured & Upper Limit \\
\mr 
Pre-\ac{O3} & 61 & 35 & 22 & 31 \\
Pre-\ac{O3} except for & & & & \\
 \ac{LHO} squeezer table & 65 & 43 & 7 & 34 \\
\br
\end{tabular}
\end{table}

The tuning procedure, which was designed primarily to provide better estimates of the coupling between a sensor and the \ac{GW} data where the coupling could not be measured due to \ac{PEM} sensor saturations, works as intended for the \ac{O3} events once the \ac{LHO} squeezer accelerometer correction is applied. Event validation experts using \textsc{PEMcheck} no longer have to investigate unrealistically large claims of environmental effects in the \ac{GW} data. Instead, this is now done automatically by the tuning process. In addition, tracking which measured values of the pre-run \ac{CF}s are consistently need tuning during events may indicate to detector commissioners that changes to the instrument made during the run may have affected the environmental coupling of the detector and that some subset of the \ac{CF}s must be re-measured.

\subsubsection{GW200115\_042309}\label{ssec:S200115j}

The \textsc{PEMcheck} analysis of data near GW200115\_042309 (hereafter GW200115) in \ac{LHO} produced the lowest single $c$-statistic of the entire dataset. This \ac{GW} event, first reported in~\cite{200115discovery}, is likely the result of a neutron star-black hole collision.

The time-frequency tile with the lowest $c$-statistic is located $\unit[0.144]{s}$ prior to the merger near $\unit[1417]{Hz}$. The \ac{PEM} channel which produced the lowest $c$-statistic tile is H1:PEM-CS\_MAG\_LVEA\_VERTEX\_XYZ\_DQ with $c=0.108.$ This channel is not recorded in \ac{aLIGO} \ac{GW} frame files but is instead calculated by the \textsc{PEMcheck} code by taking the quadrature sum of three channels, each corresponding to a different axis of the triaxial magnetometer placed near the \ac{LHO} beamsplitter which directs input light down the two interferometer arms. 

We manually investigate magnetometer and \ac{GW} data near the time-frequency tile identified by \textsc{PEMcheck} to evaluate the claim of environmental coupling. Comparing the constant-Q transforms of the strain and magnetometer channels, we find that there is some signal at that time and frequency in the magnetometer data~\cite{qtrans1,qtrans2,constQ}. However, there are other instances of magnetic signals at that frequency near the \ac{GW} time which are more energetic which do not produce any response in the \ac{DARM} channel. For this reason, we do not conclude that there is environmental contamination from the local magnetic environment in this instance. The difficulty associated with obtaining a measured magnetic coupling at high frequencies led to the environmental coupling being overestimated. The tuning procedure was not initiated in this instance; the upper limit estimate on the coupling in the region of $\unit[1417]{Hz}$ is characterized by a comb feature where the upper limit estimate of magnetic field to \ac{DARM} coupling is considerably larger in a few $\unit[]{Hz}$ band. Above $\unit[1417]{Hz}$, the coupling is tuned down, but this frequency happens to be the point in the comb that, when the projected \ac{ASD} is calculated, comes the closest to the actual background \ac{DARM} \ac{ASD} without exceeding it. Since \textsc{PEMcheck}'s results are subject to human review, the apparent overestimation of magnetic coupling at this point would have been identified during the internal event validation process.

\begin{figure}
\includegraphics[width=\textwidth]{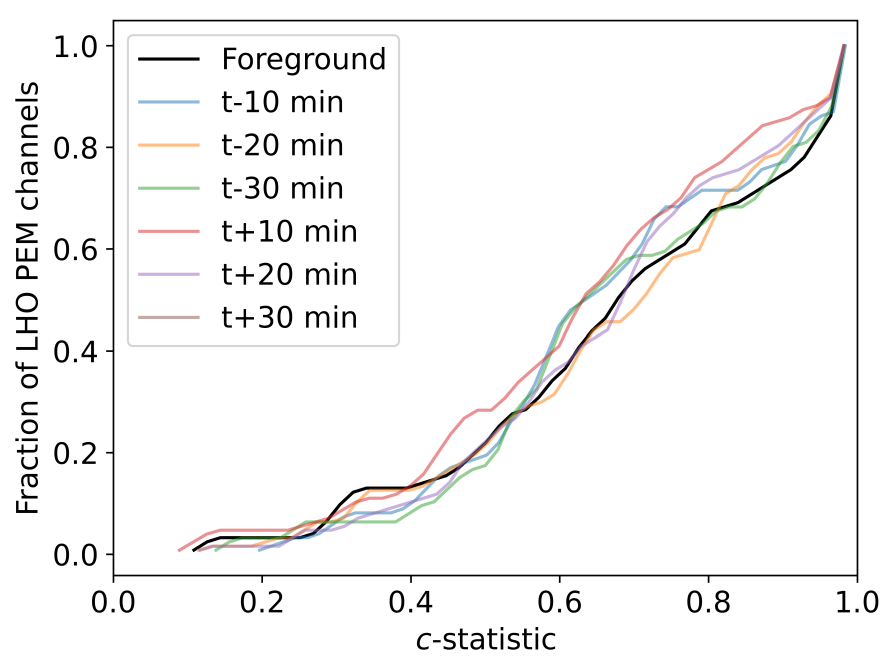}
\caption{\label{fig:shift_bbh} \ac{CDF} of the $c$-statistic for each run of \textsc{PEMcheck} at \ac{LHO} with GW200115's properties. The \ac{CDF} marked ``foreground" corresponds to \textsc{PEMcheck} run at the actual time of the \ac{GW} event.}
\end{figure}

GW200115 also stands out as having many channels report a moderate $c$-statistic. Its \ac{CDF}$(c)$ is almost always the highest of all the other traces shown in figure~\ref{fig:spaghetti_plot}. Of the sensors identified by \textsc{PEMcheck} with a $c\leq0.7$, $10$ corner station accelerometers report the highest evidence for coupling at $50$--$\unit[53]{Hz}$ at $\unit[3.64]{s}$ prior to the merger time.  This is above the $c$ threshold of $0.2$ proposed as the value of the $c$-statistic needed to initiate manual review. To confirm that these channels are not coupling to \ac{DARM}, we examine the distribution of $c$-statistics at nearby times to determine whether the collection of \ac{CS} accelerometers at near-threshold $c$ is exceptional. As illustrated in figure~\ref{fig:shift_bbh}, the distribution of channel $c$-statistics changes little when \textsc{PEMcheck} is run using the same parameters inferred in low-latency for the time-frequency track at $\pm10$, $20$ and $\unit[30]{min}$ from the actual \ac{GW} time. Like the foreground time, the distribution of moderately-valued $c$-statistics is in each instance of the \textsc{PEMcheck} computation driven by corner station accelerometers. Ground motion at the corner station is heightened due to split mini air handling units operating in the time around the \ac{GW} detection. No noise transients due to the air handling units are seen in the hour of strain data centered on the real \ac{GW} event. The lack of evidence for environmental coupling by this channels indicates that that a minimum $c$ of $0.2$ or less is likely a reasonable threshold for manual review.

\subsubsection{GW190917\_114630}\label{sssec:S190717u}

The event with the second lowest value of $\mathrm{min}(c)$ is GW190917\_114630 at \ac{LHO}. \textsc{PEMcheck} finds that the potential environmental contamination may be witnessed by the pair of triaxial magnetometers located in the electronics room at \ac{EX}. The minimum $c$-statistic tile is at $\unit[837]{Hz}$ and $\unit[0.084]{s}$ after the merger time and has $c=0.117$. Following the same argument as in section~\ref{ssec:S200115j}, the constant Q-transforms of magnetometer and \ac{GW} data do not support environmental contamination in the data. Again, poorly-constrained high-frequency magnetic coupling functions are responsible for overestimating the potential for environmental contamination in this case.

\subsubsection{GW190930\_133541}\label{sssec:S190930s}

The final event with a low value of $c$ we examine is GW190930\_133541 at \ac{LHO}. An \ac{LHO} accelerometer monitoring motion of the vacuum enclosure which houses the signal recycling cavity at $\unit[50.6]{Hz}$ and $\unit[0.352]{s}$ prior to the merger time is responsible for the $c$-statistic of $0.129$ for this \ac{GW} event~\cite{PEMpage,ligorcdesign}. This purported coupling is not supported by studying constant-Q transforms of the \ac{GW} strain and \ac{PEM} channel data near the event.  The \ac{CF} data for this sensor near the frequency identified by \textsc{PEMcheck} is set by an upper limit estimate of the coupling, which is an order of magnitude estimate rather than a more precise measure of the coupling, as described in section~\ref{ssec:envcoupling}. This could explain the discrepancy between the coupling projection and the lack of environmental coupling seen in the \ac{GW} data from around the event. There is not support for environmental coupling for this event.

\begin{figure}[h]
\includegraphics[width=\textwidth]{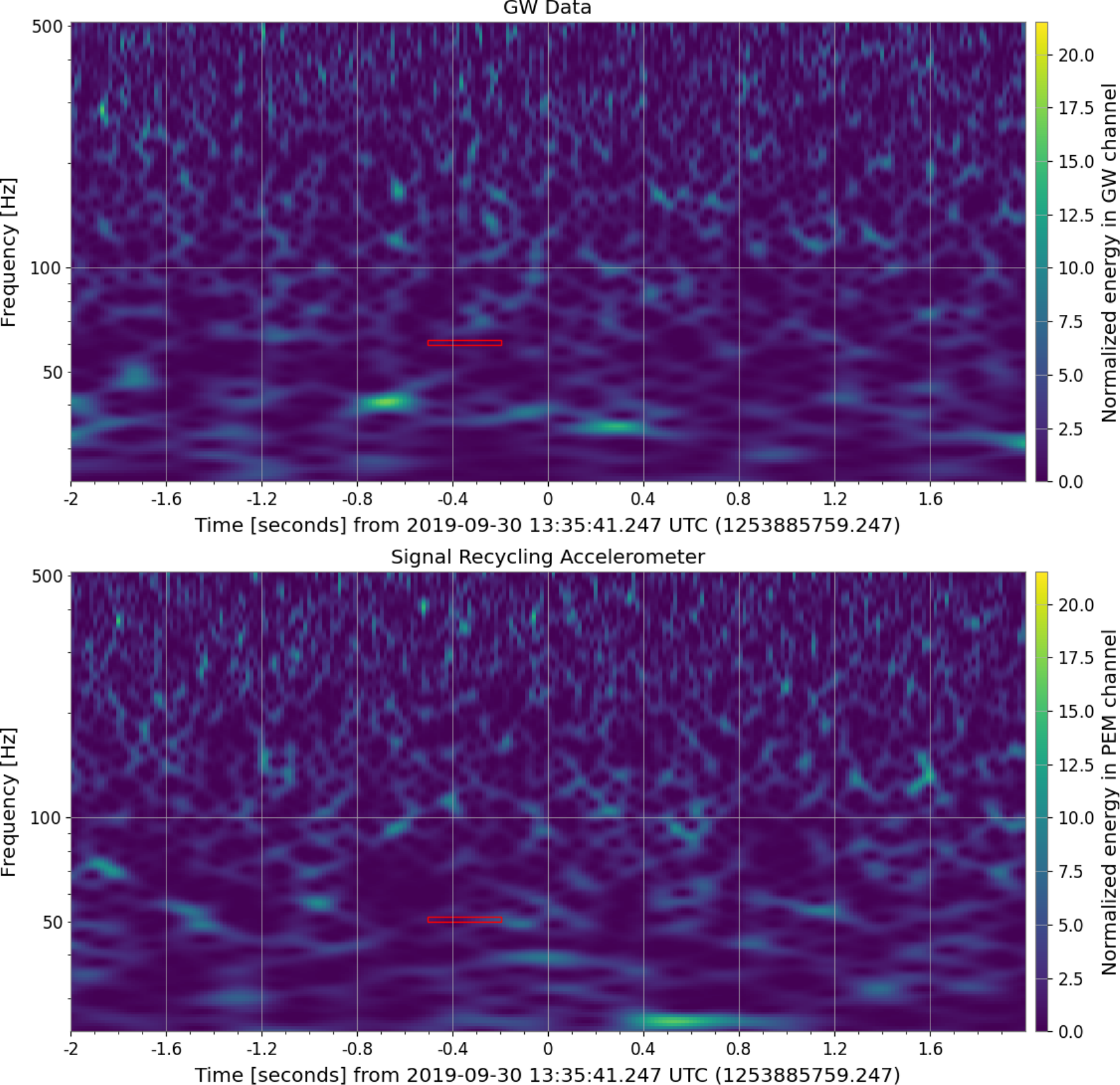}
\caption{\label{fig:190930scomp} Constant-Q transform of \ac{LHO} \ac{GW} strain (top) and signal recycling cavity beamtube accelerometer (bottom) data centered on GW190930\_133541's arrival time. A red box is overlaid on each plot showing the time-frequency window identified by \textsc{PEMcheck} as having the highest likelihood of coupling. Higher-energy transients in the accelerometer compared to the tile flagged by \textsc{PEMcheck} appear in the accelerometer data but do not correspond with transients at $\unit[50.6]{Hz}$ in the \ac{GW} data. This indicates that \ac{CF} is overestimating the coupling at this point and that the data is not subject to environmental contamination. Furthermore, the coupling estimate by \textsc{PEMcheck} is driven by the transient beginning near $t_c-\unit[0.2]{s}$, but the \ac{NR}-approximated inspiral waveform lies at a higher frequency by the time this transient appears.}
\end{figure}

While the methodology of manual vetting of \ac{GW} candidates is beyond the scope of this work, we present some arguments usually employed in the manual vetting process to establish that this event is free from environmental contamination from this \ac{PEM} sensor. We compare constant Q-transforms of \ac{PEM} sensor and \ac{GW} data from the \ac{GW} event time to look for occurrences of coincident high-energy transients in the two datastreams. These two Q-transforms are shown in figure~\ref{fig:190930scomp}. There are time-frequency tiles at $\unit[50.6]{Hz}$ in the accelerometer Q-transform with higher energy than the tile identified by \textsc{PEMcheck} as having the lowest $c$-statistic. Some examples of these transients are found at approximately $t_c-\unit[1.5]{s}$ and $t_c+\unit[1.1]{s}$. There is no corresponding response to these transients in the \ac{GW} data. These higher energy tiles were ignored by \textsc{PEMcheck} because they lie outside the calculated time-frequency track of the \ac{GW} signal. Second, we investigate the frequency-domain behavior of the signal recycling cavity beamtube accelerometer to see whether it is exceptional in some way near the time of GW190930\_133541. Longer-duration spectrograms of channel data indicate that the accelerometer is not witnessing any extraordinary environmental noise at $\unit[50.6]{Hz}$ during the event time. Lastly, the estimated ambient noise level witnessed by this sensor at this frequency is close to the \ac{GW} sensitivity when projected into \ac{GW} data~\cite{O3PEM,PEMpage}. Because of this, random fluctuations in the \ac{PEM} \ac{ASD} data should occasionally lead to claims of coupling that need to be followed up on.

\subsection{Simulated \ac{GW} Signals}\label{ssec:simresults}

We examine the performance of \textsc{PEMcheck} in an extreme scenario: one or more pipelines claim detection of a \ac{GW} during a period of high environmental contamination of the \ac{DARM} data. Thunderclaps can induce a signal in \ac{DARM} by abruptly increasing ground motion in the tens of Hertz band~\cite{O3PEM,O2O3Detchar}. An especially loud thunderclap was visible in \ac{DARM} as well as accelerometers located at \ac{LLO}'s \ac{EY} station in May of 2019~\cite{2019MayThunderAlog}. We consider the results of running \textsc{PEMcheck} on 37 hypothetical \ac{GW} signals with the properties of GW190521\_030229 (hereafter GW190521) occurring during this particular thunderclap. The properties of GW190521 used for this analysis are given in table~\ref{tab:190521g_props}.

\begin{table}[h]
\caption{\label{tab:190521g_props} Estimates of source frame masses and dimensionless spins for GW190521 taken from~\cite{190521discovery}. These parameters were used to generate the tracks in figure~\ref{fig:thunder_tracks_in_darm}. No specific $t_c$ is given, as it differs for each of the $37$ \ac{GW} tracks in the analysis.}
\begin{indented}
\item[]\begin{tabular}{@{}ll}
\br
Progenitor property & Value \\
\mr
$m_1/M_{\odot}$ & $85$\\
$m_2/M_{\odot}$ & $66$\\
$\chi_1$ & $0.69$\\
$\chi_2$ & $0.73$\\
\br
\end{tabular}
\end{indented}
\end{table}

\begin{figure}[h]
\includegraphics[width=\textwidth]{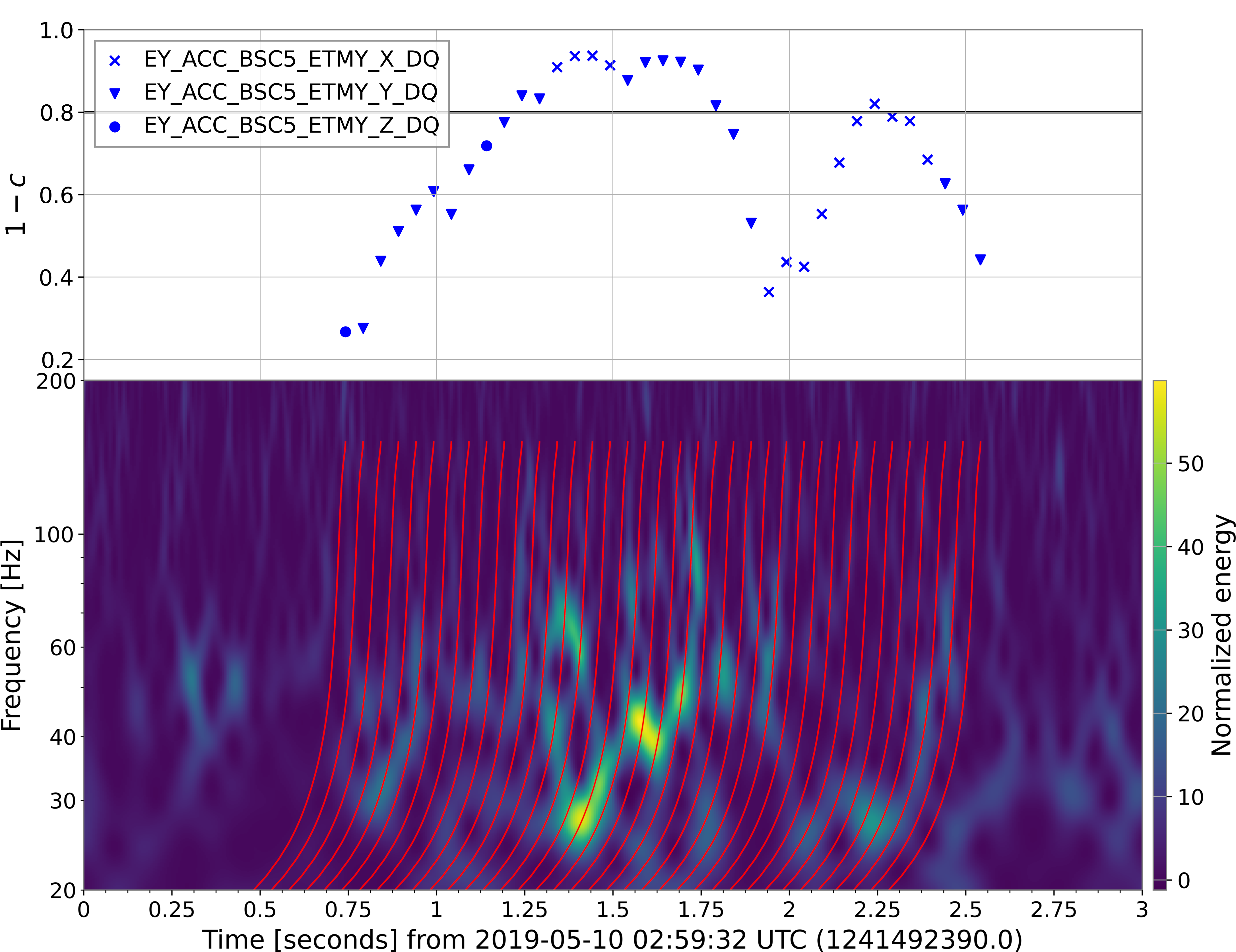}
\caption{\label{fig:thunder_tracks_in_darm} Below: constant-Q transform of \ac{LLO} \ac{GW} strain data during a thunderclap witnessed by \ac{PEM} sensors. The 37 red time-frequency tracks overlaid on the thunderclap data represent 37 hypothetical \ac{GW} time-frequency tracks with the properties given in table~\ref{tab:190521g_props} vetted by \textsc{PEMcheck}. Above: the lowest value of $c$ found by \textsc{PEMcheck} for the series of hypothetical \ac{GW} tracks overlaid on thunderclap data. The channel with the lowest $c$-statistic for a given trial is denoted by the marker style. The bold line corresponds to $c=0.2$.}
\end{figure}

We find that \textsc{PEMcheck} routinely identifies a low $c$ in the fictitious \ac{GW} signals during the thunderclap shown in figure~\ref{fig:thunder_tracks_in_darm}. The frequency identified as the likeliest one to be contaminated is $\unit[48\pm2]{Hz}$ in each trial. The tiles with the most normalized energy in the $\unit[40-50]{Hz}$ band correspond to \ac{GW} strains of $2.54\times10^{-22}$. During the thunderclap, the L1:PEM\_EY\_ACC\_BSC5\_ETMY\_Y\_DQ accelerometer recorded a maximum displacement of $\unit[1.22\times10^{-8}]{m}$. Projecting this into the \ac{GW} strain signal with the sensor's \ac{CF}, we find that predicted strain value associated with that amount of sensor motion at that frequency is $2.36\times10^{-22}.$ As a point of comparison, GW150914's strain reached a maximum amplitude of $2.63\times10^{-22}$ in the same frequency band~\cite{150914GWOSC}.

Should an event validation expert see a \ac{GW} candidate with multiple nearby sensors reporting a low $c$-statistic at similar frequencies, an appropriate recourse would be to contact environmental monitoring and noise subtraction experts for further analysis and study of the environmental noise background local to the candidate event.

\section{Conclusions}
\label{sec:conclusions}
In this work, we have demonstrated the feasibility of the \textsc{PEMcheck} tool to identify potential coupling between the \ac{aLIGO} detectors and their environment for \ac{GW} candidate events. \textsc{PEMcheck} uses coupling functions measured between \ac{PEM} sensors and the \ac{DARM} measurement to project noise witnessed by \ac{PEM} sensors into the \ac{GW} strain data stream and determine the likelihood that the \ac{GW} data at a given time and frequency is a result of environmental noise. This is quantified by the contamination (or $c$) statistic.

We have tested the performance of \textsc{PEMcheck} by analyzing the 79 \ac{GW} confident events seen by one or more of the \ac{aLIGO} detectors in \ac{O3}. This amounts to 149 individual runs of \textsc{PEMcheck}. The events with the lowest value of the contamination statistic are studied in further detail. When subject to manual inspection, we conclude that the potential environmental coupling is not borne out in each case. This is juxtaposed with a clear case of environmental coupling caused by a thunder clap. From the results of running \textsc{PEMcheck} on real and simulated \ac{GW} signals a criterion of $c\leq0.2$ is suggested as the standard for initiating human review of \textsc{PEMcheck} results by event validation experts.

We show that \textsc{PEMcheck} is capable of warning event validation experts if there is substantial environmental coupling. We find that the tuning procedure used by \textsc{PEMcheck} provides a rough estimate of the environmental coupling in real-time, reducing the demand for hand-vetting of events by environmental noise experts. However, claims of environmental coupling continue to be checked by hand in the case of a high likelihood of environmental contamination or detections of \ac{GW}s from novel sources. \textsc{PEMcheck} is being used in the \ac{DQR} to study environmental noise contamination in every \ac{GW} observed by that \ac{aLIGO} detectors in \ac{O4}.

Future directions for development may include closer integration with utilities for characterizing changes in sensor performance over time~\cite{LIGOcam} and designing a framework for constraining coupling from \ac{PEM} sensors using loud transients seen in those channels and not in \ac{DARM} to further tune estimates of the coupling function, similar to the approach demonstrated in section~\ref{sssec:S190930s} partially manually vetting GW190930\_133541. The results of the tuning procedure should be tracked to motivate injected campaigns if \ac{CF}s tend to become inaccurate as observing runs continue, especially in cases where upper limit estimates consistently need adjusting. Nonlinear coupling is not accounted for by current \ac{CF}s and thus \textsc{PEMcheck}, but would be an important development for the future. A more rigorous approach to trials factors may improve the interpretability of numerical results. The limitations of this procedure, e.g. in \ac{PEM} sensor coverage, will inform the designs for future GW detectors, such as Einstein Telescope and Cosmic Explorer, for which environmental noise will continue to be crucial to confront~\cite{ETwhitepaper,CEwhitepaper}. Since the \textsc{PEMcheck} analysis requires \ac{CF}s, sensor and \ac{DARM} data, the same procedure used to search for evidence of environmental coupling could be used in these future \ac{GW} detectors.

\ack
The authors are grateful to Ronaldas Macas and Jane Glanzer for their careful review of the \textsc{PEMcheck} code and manuscript, respectively. We also thank Derek Davis for supporting \textsc{PEMcheck}'s integration with the \ac{O4} \ac{DQR}.

The authors acknowledge support from National Science Foundation Grants PHY-2207535 and PHY-2207713. The authors are grateful for computational resources provided by the LIGO Laboratory and supported by National Science Foundation Grants PHY-0757058 and PHY-0823459. 

This research has made use of data or software obtained from the Gravitational Wave Open Science Center (gwosc.org), a service of LIGO Laboratory, the LIGO Scientific Collaboration, the Virgo Collaboration, and KAGRA. LIGO Laboratory and Advanced LIGO are funded by the United States National Science Foundation (NSF) as well as the Science and Technology Facilities Council (STFC) of the United Kingdom, the Max-Planck-Society (MPS), and the State of Niedersachsen/Germany for support of the construction of Advanced LIGO and construction and operation of the GEO600 detector. Additional support for Advanced LIGO was provided by the Australian Research Council. Virgo is funded, through the European Gravitational Observatory (EGO), by the French Centre National de Recherche Scientifique (CNRS), the Italian Istituto Nazionale di Fisica Nucleare (INFN) and the Dutch Nikhef, with contributions by institutions from Belgium, Germany, Greece, Hungary, Ireland, Japan, Monaco, Poland, Portugal, Spain. KAGRA is supported by Ministry of Education, Culture, Sports, Science and Technology (MEXT), Japan Society for the Promotion of Science (JSPS) in Japan; National Research Foundation (NRF) and Ministry of Science and ICT (MSIT) in Korea; Academia Sinica (AS) and National Science and Technology Council (NSTC) in Taiwan.

This publication has been assigned LIGO document number P2300254.

\section*{References}
\bibliography{references}

\end{document}